\documentclass[prd,twocolumn,preprintnumbers,nofootinbib]{revtex4}
\usepackage{graphicx}
\usepackage{color}
\usepackage{float}
\newcommand{\rthis}[1]{\textcolor{black}{#1}}
\usepackage{amsfonts,amsmath,amssymb}
\usepackage[plainpages=false, colorlinks=true, anchorcolor=blue, linkcolor=blue, citecolor=blue, bookmarks=false]{hyperref}
\usepackage{natbib}
\usepackage{enumitem}
\usepackage{lipsum}
\usepackage{multirow}
\usepackage{graphicx}
\usepackage{array}
\pdfoutput=1
\begin{document}
%\markboth{Authors' Names}
%\newcommand{\mthis}[1]{\textcolor{red}{#1}}
\newcommand{\bthis}[1]{\textcolor{black}{#1}}
\newcommand{\apjl}{Astrophys. J. Lett.}
\newcommand{\apjs}{Astrophys. J. Suppl. Ser.}
\newcommand{\aap}{Astron. \& Astrophys.}
\newcommand{\nar}{New  Astronomy Reviews}

\newcommand{\aj}{Astron. J.}
\newcommand{\araa}{Ann. Rev. Astron. Astrophys. } %ARA$\&$A}
\newcommand{\mnras}{Mon. Not. R. Astron. Soc.}
\newcommand{\ssr}{Space Science Revs.}
\newcommand{\apss}{Astrophysics \& Space Sciences}
\newcommand{\jcap}{JCAP}
\newcommand{\pasj}{PASJ}
\newcommand{\pasp}{PASP}
\newcommand{\pasa}{Pub. Astro. Soc. Aust.}
\newcommand{\physrep}{Phys. Rep.}

\title{Characterization of the GRB prompt fundamental plane using Fermi-GBM data}
\author{S. \surname{Pradyumna}}\altaffiliation{E-mail:ep18btech11015@iith.ac.in}

\author{Shantanu  \surname{Desai}}  
\altaffiliation{E-mail: shntn05@gmail.com}

\begin{abstract}
A recent work has found a tight global relation between the GRB prompt fluence, peak flux (based on the optimum time scale determined from the Bayesian Blocks-based analysis)  and duration using data from the Fermi Gamma-Ray burst monitor, which they have dubbed as ``Fundamental Plane''. We quantitatively characterize the tightness of this Fundamental Plane relation by calculating the scatter in dex.   We also check for a fundamental plane using the peak flux over time scales of  64 ms, 256 ms and 1024 ms.
For our analysis, we incorporate the uncertainties in the above observables and carried out both a PCA as well as regression-based analysis.  We find that the scatter in the fundamental plane is 0.16-0.17 dex. 
\end{abstract}

\affiliation{Department of Physics, Indian Institute of Technology, Hyderabad, Telangana-502284, India}
\maketitle

\section{Introduction}
Gamma-ray bursts (GRBs) are short-duration single-shot transients with energies in the keV-GeV energy regime~\cite{Kumar}. They were first discovered in the 1970s~\cite{firstgrb}. Based on their duration ($T_{90}$), they have been traditionally divided into categories:  short and long, depending on whether $T_{90}$ is less than or greater than two seconds~\cite{Kouv93}. Long-duration GRBs have been associated with core-collapse supernova~\cite{Bloom} and short GRBs with binary neutron star mergers~\cite{Nakar}. GRBs are located at cosmological distances, although a distinct cosmological signature in the GRB light curves is yet to be unequivocally  demonstrated~\cite{Singh}.

However, this distinction into two categories and the physical association of long and short GRBs to core collapse SN/compact object mergers is not completely sacrosanct. For more than three decades, claims have been made for a third category of GRBs~\cite{Horvath98,Kulkarni,Bhave} (and references therein). Also, exceptions to the above association  of long and short GRBs to core collapse SN and  compact object mergers have also been found~\cite{Zhang,Amati20,Ahumada,Rossi21}.

\rthis{Over the past two decades, a large number of correlations between two or more GRB observables in both the prompt and afterglow emission phase have been found~\cite{DainottiAmati,Ito}. We recap  these GRB fundamental plane relations found in literature. A discussion of the bivariate correlations can be found in the above reviews, and will not be discussed here.
The first fundamental plane relation between the isotropic energy, peak energy in the $\nu F_{\nu}$ spectrum, and the break-time of optical afterglow light curve was found in ~\cite{Zhang05} with negligible dispersion. A fundamental plane relation between the peak energy, peak luminosity, and the luminosity time (defined as the ratio of isotropic equivalent energy and peak luminosity, and hence implicitly depends on T90) was also found with a scatter of between 13-20\%~\cite{Tsutsui}.} 

\rthis{Fundamental plane relations have also been discovered using the afterglow parameters. Using X-ray afterglows of long GRBs with an extended plateau phase,   a tight fundamental plane relation between the rest frame end time of the plateau, X-ray luminosity, and the peak luminosity in the prompt emission phase was found, with an intrinsic scatter of 27\%~\cite{Dainotti16}. This Dainotti relation was further affirmed using the Gold sample of 45 SWIFT-detected GRBs with a scatter of 30\%~\cite{Dainotti17} and the platinum sample of long GRBs with an intrinsic scatter of 22\%~\cite{Dainotti20}.}

Most recently,~\citet{Zhang22} (Z22 hereafter)  revisited the problem of GRB classification using Fermi-GBM data  to determine the optimum variables, which could be used for such a classification. They found a tight relation between the GRB fluence  in the 10-1000 keV range ($f$), peak flux ($F$) and prompt duration ($T_{90}$) in logarithmic space, given by:
\begin{equation}
    \log f = 0.75 \log  T_{90} + 0.92 \log F - 7.14 
    \label{eq:zhang}
\end{equation}
where $\log$ refers to logarithm to base 10.
Here, T90  and the GRB fluence were obtained from the online public Fermi-GBM catalog~\cite{Gruber,GBM,fermi14,vonKienlin20}.
The peak flux was calculated by Z22 on the timescale obtained by the Bayesian Block algorithm~\cite{Scargle13}. Thereafter, a new data vector was constructed from the first two principle components and GRB classification using Gaussian mixture model (GMM) was implemented on this data vector. Linear discriminant analysis was then applied to find the optimum boundary between the two GMM generated components.
This combination of peak,flux, fluence and duration was also observed to differentiate between short and long GRBs with much more efficacy than only T90.

In this work, we try to quantify the scatter in the aforementioned fundamental plane using the above variables to determine how tight is the fundamental plane. We also try three other variants of the GRB flux, obtained from the online Fermi-GBM catalog. For these analyses, we  incorporate the errors in T90, fluence as well as peak flux to scrutinize the robustness of the fundamental plane. We use the same methods as in our recent work, where we characterized the fundamental plane consisting of  the temperature, gas mass and scale radius of the X-COP cluster sample~\cite{Pradyumna22}.

This manuscript is structured as follows.  We describe the data and methodology used for our analysis in Sect.~\ref{sec:data}. Our results are discussed in Sect.~\ref{sec:results}. We conclude in Sect.~\ref{sec:conclusions}.

\section{Data and Methods}
\label{sec:data}
The characterization of fundamental plane defined in  Eq.~\ref{eq:zhang} requires measurements of three quantities - ($f$, $T_{90}$, $F$).
The fluence ($f$) is the defined as the  total energy received per unit area of the detector from a GRB in $10 - 1000$ keV range.
$T_{90}$ is defined as the time elapsed between collection of 5\% of the total fluence to 95\% of the total fluence. Peak flux ($F$) is the maximum flux of photons detected in a given timescale. 
In this work, data from the fourth release of Fermi GRB was used \cite{Gruber,GBM,fermi14,vonKienlin20}.   The data release\footnote{The data can be found at: \url{https://heasarc.gsfc.nasa.gov/W3Browse/fermi/fermigbrst.html}} contains a catalog of  3222 GRBs detected  from 12th July 2008 until 11 February 2022.  For each GRB, T90, $f$ and $F$ are provided~\citep{vonKienlin20}, where both $f$ and $F$ have been computed  in two different energy ranges: 10-1000 keV and 50-300 keV. The peak fluxes are computed for three different timescales: 64 ms, 256 ms, as well as 1024 ms. We denote each of the above fluxes as $F_{64}$, $F_{256}$, and $F_{1024}$. For each of these variables, the 1$\sigma$ error bars have also been provided by the Fermi-GBM collaboration. More details  on the estimation of these quantities can be found in ~\cite{vonKienlin20}. For our analysis, we have used the fluxes and fluences in the 10-1000 keV energy range to emulate the analysis in Z22.  Note that the online Fermi-GBM catalog does not provide the peak flux based on the time scale provided by the Bayesian block method ($F_\mathrm{BB}$). This flux was provided to us by the authors of Z22 (private communication) for 2989 out of 3222 GRBs.
For estimating the best fit plane using $F_{64}, F_{256}$ and $F_{1024}$, we include all the 3222 GRBs. On the other hand, because $F_\mathrm{BB}$ is available only for 2989 GRBs, the best fit plane was obtained for this subset of 2989 GRBs. 

% please mention the median fractional error in T90 and all the 3 fluxes.
While the median fractional error in $T_{90}$ is found to be 12\%, the mean fractional error is about 42\%, indicating the presence of outliers. The median fractional error in $f, F_{64}, F_{256}$, and $F_{1024}$ is equal to  1.6\%, 19\%, 11\% and 7\%, respectively. No significant deviations between the median and mean fractional errors were observed in the case of quantities other than $T_{90}$. $F_\mathrm{BB}$ estimated in Z22 does not have uncertainties provided. So we did the fundamental plane analysis for this flux   in two ways: without assuming any error bars as well as by assuming an uncertainty of 10\%, which is comparable to that for the other fluxes.

In this work, the fundamental plane analysis was carried out in two ways: (1) PCA analysis using  the generation of mock data sets based on the observed uncertainties (2) by Bayesian regression analysis.  We estimated the scatter in the fundamental plane  for both these analyses. We now describe each of these analyses in turn.

\subsection{PCA analysis}
We use principal component analysis (PCA) to fit a plane to the three variables ($T_{90}, F, f$) in logarithmic space. PCA in three dimensions returns three vectors $P_1, P_2$, and $P_3$, which are known as principal components. These principal components are such that they are orthogonal to each other and arranged in decreasing order based on contribution to the total variance. Therefore, the third component $P_3$ can be considered as a normal to the best fit plane of the data.
This method has been extensively used in dimensionality reduction.

Similar to our recent work on XCOP sample~\cite{Pradyumna22} (see also ~\cite{Fujita}), we incorporated the errors in the GRB variables in the PCA analysis by constructing synthetic datasets and applying PCA to each of these mock datasets. By assuming the uncertainties on the quantities to be Gaussian, simulated data sets were generated by choosing randomly from a normal distribution with the standard deviation equal to the observational uncertainties. Since we are applying PCA in logarithmic space, the uncertainty is obtained by error propagation. For example, the uncertainty in $\log T_{90}$ is given by: $\sigma_{\log (T_{90})} = \frac{\sigma_{T_{90}}}{\ln(10) \, T_{90}}$, and so on for the other two variables.

The third component $P_3$ returned by PCA is a vector $(a, b, c)$ normal to the best-fit plane, such that we obtain a plane equation of the form:
\begin{equation}
    a\log_{10} T_{90} + b \log_{10} F + c \log_{10} f = 0
    \label{eq:PCA}
\end{equation}
where quantities are scaled by their mean values to center of the plane at origin. Similar to ~\cite{Pradyumna22}, the scatter was calculated as half the difference between $84^\mathrm{th}$ and $16^\mathrm{th}$ quantiles of orthogonal distances of data from the best fit plane.

\subsection{Bayesian regression}
As an alternative to standard PCA, we also did a regression analysis in logarithmic space between the peak flux and fluences to characterize the scatter of the fundamental plane. More specifically we fit the data to the following regression relation:
\begin{equation}
    \log_{10} f = \alpha \log_{10} F + \beta \log_{10} T_{90} + \delta
    \label{eq:MCMC}
\end{equation}
From Eqs.~\ref{eq:MCMC} and \ref{eq:PCA}, we see that $\alpha = -b/c$ and $\beta  = -a/c$.
We obtain $\alpha$ and $\beta$ using the MCMC sampler {\tt emcee}~\cite{emcee} from the following \rthis{Gaussian} likelihood:
\begin{equation}
    -2 \ln L = \sum_i (2 \pi \sigma_i^2) + \sum_i (d_i/\sigma_i)^2, 
    \label{eq:likelihood}
\end{equation}
where summation is over all data, with $d_i$ being the distance of data from the plane given as:
\begin{equation}
    d_i = \frac{| \log_{10} f - (\alpha \log_{10} F + \beta \log_{10} T_{90}  + \delta)| }{\sqrt{\alpha^2 + \beta^2 + 1}}.
\end{equation}
and $\sigma_i^2 = \alpha^2 (\sigma_{\log F_i})^2 + \beta^2 (\sigma_{\log T_{90, i}})^2 + (\sigma_{\log f_i})^2 + \sigma_{\mathrm{int}}^2$. Here, $\sigma_\mathrm{int}$ is the internal scatter of the best fit relation about the data. 
The priors for $\alpha, \beta$ and $\delta$ were chosen as $(0,5), (0,5)$ and $(-10, 10)$, based on the best-fit values obtained using PCA. 
We then characterize the scatter of the fundamental plane using the same method as in our analysis of radial acceleration relation~\citep{PradyumnaRAR}.

\section{Results}
\label{sec:results}
\begin{figure*}
    \centering
    \includegraphics[width = 2\columnwidth]{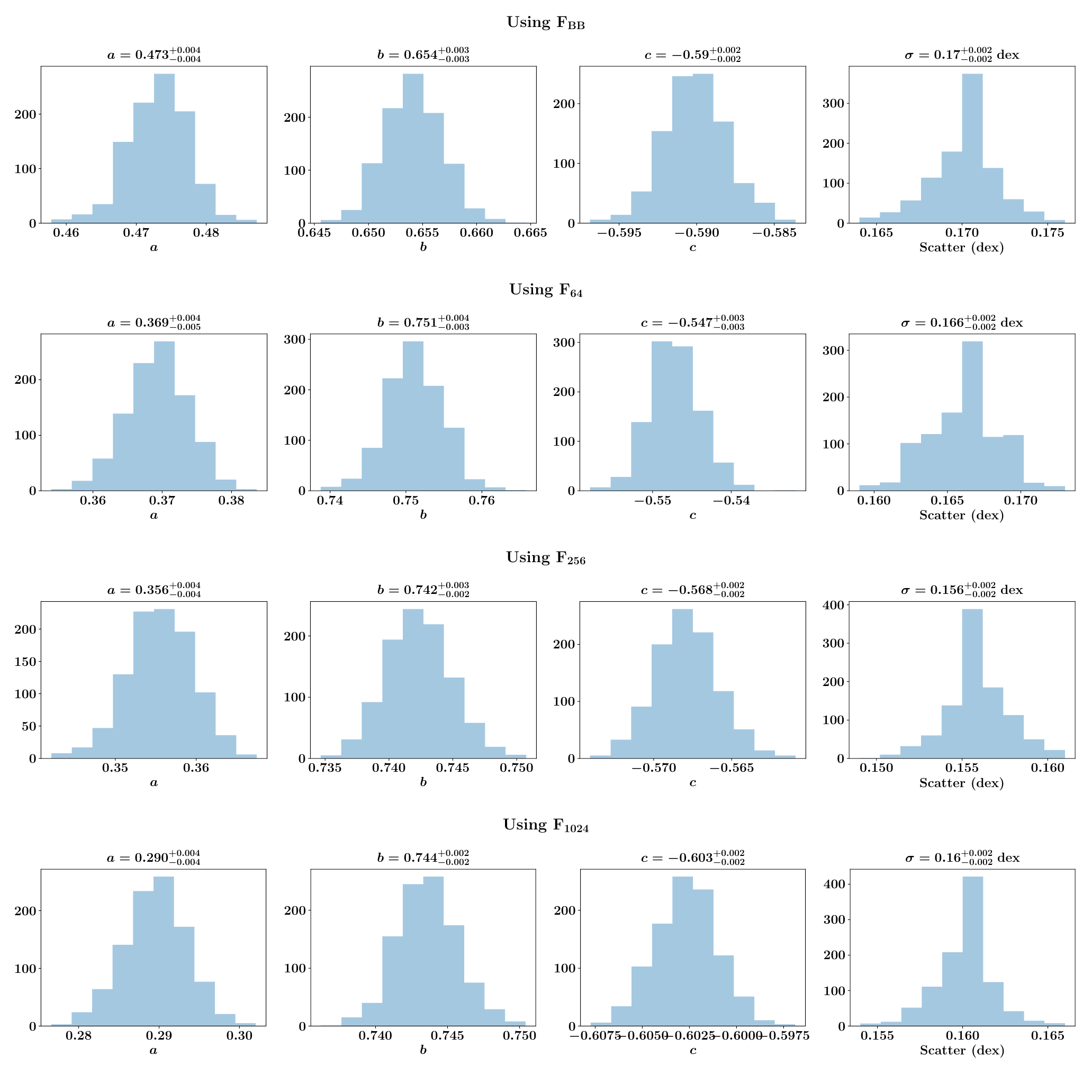}
    \caption{The Fundamental Plane parameters  along with intrinsic scatter obtained by applying PCA to synthetic data sets constructed from different combinations of fluence, fluxes and T90.}
    \label{fig:GRB_Sy}
\end{figure*}

\begin{figure*}      
    \centering  
    \includegraphics[width = 1.8\columnwidth]{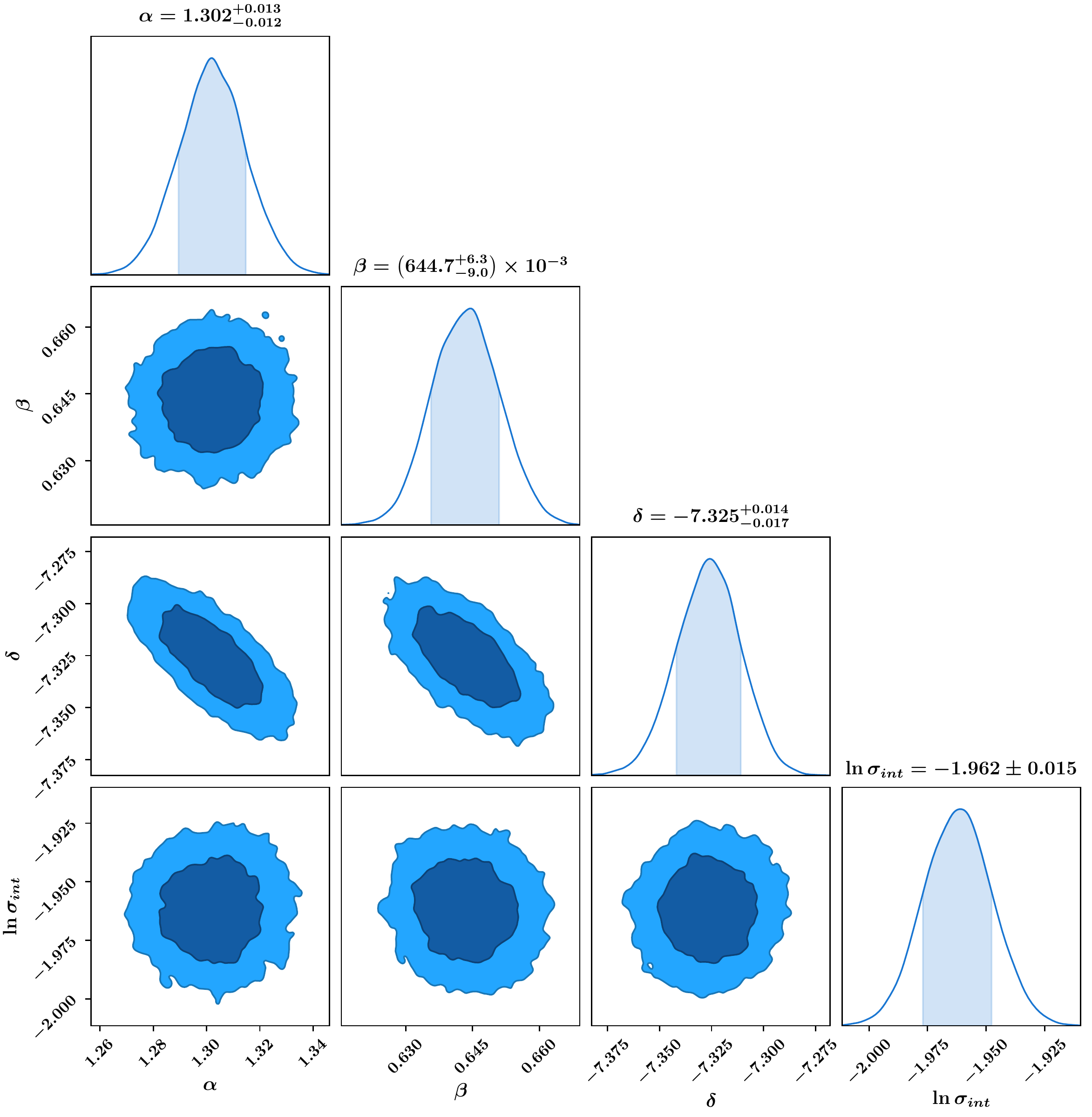}
    \caption{$68\%$ and $95\%$ credible interval plots, obtained for a relation of the type $f =  F^\alpha T_{90}^\beta$. This plot is obtained using $F_{256}$. {\tt Chainconsumer} tool \protect\cite{Chainconsumer} was employed to generate this plot}
    \label{fig:GRB_MCMC}
\end{figure*}

We carry out the PCA based analysis and the regression analysis on four different triplets of datasets in logarithmic space ($T_{90},f,F_{64}$), ($T_{90},f,F_{256}$), ($T_{90},f,F_{1024}$), and ($T_{90},f,F_\mathrm{BB}$).
For $F_\mathrm{BB}$, we carried out the analyses assuming no errors as well as assuming 10\% fractional error. The histograms of the best-fit parameters as well as the intrinsic scatter can be found  in Fig.~\ref{fig:GRB_Sy}. We also show the credible intervals obtained from regression analysis for one of the three fluxes ($F_{256}$) in Fig.~\ref{fig:GRB_MCMC}. \rthis{The marginalized contours in Fig.~\ref{fig:GRB_MCMC} also provide insight on the possible degeneracies between the different parameters. We find that $\delta$ is anti-correlated with $\alpha$ and $\beta$, but $\alpha$ and $\beta$ are uncorrelated.}
A tabular summary of all our results can be found in Table~\ref{tab:results}. The best-fit parameters and scatter mostly agree with that obtained with regression analysis, although the values disagree at 1$\sigma$ \rthis{for some of the parameters}. We find that the intrinsic scatter using PCA for $F_{64}$, $F_{256}$, and $F_{1024}$ is equal to 0.166, 0.156, and 0.16, respectively. The corresponding intrinsic scatter for $F_\mathrm{BB}$ is equal to 0.168 (without including errors in $F_\mathrm{BB}$) and 0.17 after including errors in $F_\mathrm{BB}$.   Therefore, we conclude that the scatter between the fluence, peak flux (including that obtained using Bayesian Blocks algorithm) and duration is not negligible.  We find that the scatter  using $F_{64}$, $F_{256}$, and $F_{1024}$ is comparable to that obtained using $F_\mathrm{BB}$.  A comparison of the scatter  obtained from the GRB fundamental plane in this work with other fundamental plane or scaling relations in astrophysics literature can be found in Table~\ref{table2}. As we can see,  the scatter is larger than some of the other well known scaling  or Fundamental plane relations such as the Faber-Jackson relation, Fundamental plane for elliptical galaxies, Radial acceleration relation, etc.
\rthis{However, from this table, we also see that  the scatter is comparable to the other GRB fundamental plane  relations obtained using the prompt/afterglow emission parameters.}

%\section{Discussion}
%\label{sec:discussion}

\begin{table*}[]
\centering

\begin{tabular}{|c|c|c|c|c|c|}
\hline\hline
Value of $F$ taken                 & Method & \multicolumn{1}{c|}{$a$} & \multicolumn{1}{c|}{$b$} & \multicolumn{1}{c|}{$c$} & \multicolumn{1}{c|}{Scatter (dex)} \\
\hline 
\hline
Z22\cite{Zhang22} & PCA & 0.483 & 0.593 & -0.644 & - \\
\hline
\multirow{2}{*}{$F_\mathrm{BB}$} & PCA    & $0.473 \pm 0.004$       & $0.654 \pm 0.003$       & $-0.590 \pm 0.002$      & $0.170\pm0.002$             \\
                                 & MCMC   &  $0.471 \pm 0.006$    & $0.653 \pm 0.009$        & $-0.593\pm 0.003$   &  $0.160 \pm 0.001$                           \\
\hline
\multirow{2}{*}{$F_\mathrm{BB}$ (zero uncertainty)}       & PCA & $0.472 \pm 0.005$ & $0.656 \pm 0.003$ & $-0.589 \pm 0.002$ & $0.168 \pm 0.002$ \\
                                 & MCMC   &    $0.473 \pm 0.006$ & $0.649\pm0.008$ &    $-0.596 \pm 0.003$ &    $ 0.157 \pm 0.001$\\
\hline
\multirow{2}{*}{$F_{64}$}        & PCA    & $0.369^{+0.004}_{-0.005}$ & $0.654 \pm 0.003$ & $-0.547 \pm 0.003$    & $0.166 \pm 0.002$    \\
                                 & MCMC   &    $0.363 \pm 0.005$ & $0.765\pm0.009$ &    $-0.533 \pm 0.003$ &    $ 0.168 \pm 0.002$\\
\hline
\multirow{2}{*}{$F_{256}$}       & PCA    & $0.356 \pm 0.004$ & $0.742^{+0.003}_{-0.002}$ & $-0.568 \pm 0.002$ & $0.156 \pm 0.002$ \\
                                 & MCMC   & $0.365 \pm 0.005$ & $0.738 \pm 0.008$ & $-0.567 \pm 0.003$ &$0.160 \pm 0.002$\\
\hline
\multirow{2}{*}{$F_{1024}$}      & PCA    & $0.290 \pm 0.004$ & $0.744 \pm 0.002$ & $-0.603 \pm 0.002$ & $0.160 \pm 0.002$\\
                                 & MCMC   & $0.313 \pm 0.005$ & $0.736 \pm 0.008$ & $-0.600 \pm 0.003$ & $0.162 \pm  0.002$ \\
\hline
\end{tabular}%
\caption{Summary of our PCA and regression analysis of fundamental plane  using fluence, four different estimates of peak flux, and prompt duration. We show values for $a$, $b$, $c$ (defined in Eq.~\ref{eq:PCA}) along with the scatter in dex. }
\label{tab:results}
\end{table*}

\begin{table*}
\centering
\begin{tabular}{|c|c|c|}
\hline
\textbf{Relation} & \textbf{Scatter} & \textbf{Ref.} \\ \hline
GRB $(f, F, T90)$ & 0.156-0.17 dex & This work \\ Galaxy Cluster $(T_x, m_s, r_s)$ &  $0.017-0.039$ dex & \cite{Pradyumna22}\\
GRB $(T_a, L_a,L_{peak})$ & 22-30\% & \cite{Dainotti16,Dainotti17,Dainotti20} \\
GRB $(E_{iso}, E_{peak}, t_b)$ & NA\footnote{The actual dispersion of this relation is negligible~\cite{Zhang05}. However, we could not find a  quantitative estimate of this scatter in literature.} & \cite{Zhang05} \\
GRB ($L_p,E_p, t_L$) & 13-20\% & \cite{Tsutsui} \\
RAR (SPARC)   & 0.11 dex & \cite{Mcgaugh16} \\
RAR (clusters) & 0.11-0.14 dex &  \cite{PradyumnaRAR} \\
F-J relation (ellipticals) & 0.09 dex  & ~\cite{Faber} \\
Baryonic F-J relation (clusters) & 0.07 dex & \cite{Tian21} \\
Baryonic T-F relation & 0.10 dex  & \cite{Lelli} \\
Early type galaxies ($R_e,\sigma, I_e$)  & 0.05 dex & 
\cite{Bernardi} \\
Galaxies ($Y_e, S_{0.5}^2,I_e, R_e$) & 0.05 dex &  \cite{Sanchez} \\
Black Hole $M-\sigma$ & 0.3 dex  & \cite{Gebhardt} \\
\hline
\end{tabular}
\caption{A comparison of the scatter obtained for GRB fundamental plane with that for other scaling or fundamental plane relations in Astrophysics literature. \rthis{Also shown is a comparison with other GRB related fundamental plane relations.} The acronyms RAR, F-J, T-F stand for Radial Acceleration Relation, Faber-Jackson, Tully-Fisher respectively. Note that the scatter includes observational uncertainties.}
\label{table2}
\end{table*}

\section{Conclusions}
\label{sec:conclusions}
In a recent work, Z22 argued (using the Fermi-GBM dataset) that there is a tight relation between the GRB peak flux, fluence and duration (T90) in logarithmic space, which they dubbed as the fundamental plane. The peak flux was obtained on a time scale derived by the Bayesian Block method. Based on this correlation,  a linear combination of the aforementioned variables was used for classification and this combination  was shown to be more  robust in GRB classification than T90.

Here, we quantify the tightness of this fundamental plane by calculating its scatter using the same procedure as our recent work on fundamental plane on XCOP galaxy cluster sample~\cite{PradyumnaRAR}. We also incorporate the uncertainties in the above observables. 
The fundamental plane analysis was done using both PCA based decomposition as well as a regression analysis between the flux, fluence, and duration.
We also examine the fundamental plane for three other fluxes provided by the Fermi-GBM collaboration on timescales of 64 ms, 256 ms, and 1024 ms. 

Our results are summarized in Table~\ref{tab:results}. A comparison of the scatter found in this analysis with other scaling and fundamental plane relations in literature can be found in Table~\ref{table2}. We find that the scatter in the fundamental plane varies  between 0.16-0.17 dex. We also find that the scatter using the other three fluxes is comparable to the flux obtained on the Bayesian block estimated time scale, which was used in Z22. These scatters  are higher than some of the other tight scaling relations found in astrophysics literature summarized in Table~\ref{table2}.

\rthis{Note, however that this is the first fundamental plane relation discovered in literature using GRB observables in the prompt emission phase and without invoking the GRB redshift.  Its scatter is also comparable to or better than some of  the other GRB-related fundamental plane correlations (cf. Table~\ref{table2}). Z22 used this relation for robust  classification of GRBs. However, it would interesting to see if this relation can be applied either on its own or  in conjunction with other bivariate or fundamental plane relations to parameter estimation in Cosmology.} 

\begin{acknowledgements}
We are grateful to Shuai Zhang for sharing with us the peak flux calculated using the Bayesian block algorithm, used in Z22.
\end{acknowledgements}

\bibliography{references}

\end{document}